\documentclass[journal,onecolumn,12pt]{IEEEtran}

\usepackage{cite}
\usepackage{amsmath,amssymb,amsfonts}
\usepackage{algorithmic}
\usepackage{graphicx}
\usepackage{textcomp}
\def\BibTeX{{\rm B\kern-.05em{\sc i\kern-.025em b}\kern-.08em
    T\kern-.1667em\lower.7ex\hbox{E}\kern-.125emX}}

\usepackage{subfigure}
\usepackage{multirow}
\usepackage{amsfonts}
\usepackage{balance}

\usepackage{epstopdf}

\AppendGraphicsExtensions{.gif}

\begin{document}

\title{Probabilistic Resistive Switching Device modeling based on Markov Jump processes}

\author{Vasileios Ntinas,
        Antonio~Rubio,
        Georgios Ch. Sirakoulis%
\thanks{V. Ntinas, and G. Ch. Sirakoulis are with the Department of Electrical and Computer Engineering, Democritus University of Thrace, Xanthi, Greece (e-mail: \{vntinas, gsirak\}@ee.duth.gr).} 
\thanks{A. Rubio and V. Ntinas, also, are with the Department of Electronics Engineering, Universitat Polit\'ecnica de Catalunya, Barcelona, Spain (e-mail: \{antonio.rubio, vasileios.ntinas\}@upc.edu).}}

\maketitle

\begin{abstract}
In this work, a versatile mathematical framework for multi-state probabilistic modeling of Resistive Switching (RS) devices is proposed for the first time. The mathematical formulation of memristor and Markov jump processes are combined and, by using the notion of master equations for finite-states, the inherent probabilistic time-evolution of RS devices is sufficiently modeled. In particular, the methodology is generic enough and can be applied for $N$ states; as a proof of concept, the proposed framework is further stressed for both two-state RS paradigm, namely $N=2$, and multi-state devices, namely $N=4$. The presented I-V results demonstrate in a qualitative and quantitative manner, adequate matching with other modeling approaches.
\end{abstract}

\begin{IEEEkeywords}
ReRAM devices, Probabilistic Modeling, Markov Processes, Cycle-to-Cycle variability.
\end{IEEEkeywords}

\maketitle

\section{Introduction}

Novel nano-devices with resistive switching (RS) properties have attracted a lot of attention, mainly due to their unprecedented electrical characteristics suitable for novel memory and processing systems \cite{ielmini2018memory}. At the same time, they flourish in the area of artificial neural networks and neuromorphic computing because of their adaptive conductivity enabling them as the electronic analog to the biological synapses. Fast switching, non-volatility and low energy consumption are just a few of their evident advantages. Devices like resistive RAM \cite{wong2012metal}, phase-change memories \cite{fong2017phase}, and ferroelectric RAM \cite{chanthbouala2012ferroelectric} can modify their resistivity (resistive switching) under the application of external electrical stimulus, while they maintain it unchanged in the lack of stimulus.

Since 2008, when HP Labs \cite{strukov2008missing} managed to bridge RS devices with the memristor conception \cite{chua1971memristor}, a wide range of mathematical models have been developed to describe the behavior of the aforementioned fabricated devices \cite{biolek2009spice,kvatinsky2012team,ascoli2013memristor,jiang2016compact,ntinas2018complete,messaris2018data}. From atomistic-level precision up to more abstract-level compact models, all models aim to provide the necessary tools for the researchers as well as circuit and system designers (in case of compact models) to understand and utilize the novel characteristics of RS devices \cite{ielmini2017physics} targeting on novel RS related applications like memories and in memory computing. 
More specifically, the existing compact models are able to either describe the underlining physical mechanisms or, phenomenologically, the behavior of the devices. However, RS is governed by stochastic phenomena at microscopic level \cite{jo2009programmable} that most of the compact models lack to capture, thus, they insert random perturbations to the state variables or parameters to incorporate such behavior \cite{li2015variation}. On the other hand, another category of inherently stochastic models is the kinetic Monte Carlo (kMC) based modeling approach \cite{yu2011stochastic,menzel2015understanding}, which describes at microscopic level the field-enabled stochastic dispositions of ions within the switching layer. However, kMC is not a compact modeling approach, while there are only a few compact models developed on stochastic approaches of RS limited to the two-state (binary) switching of the device \cite{dowling2020probabilistic}.

In this work, for the first time, a versatile mathematical framework for the probabilistic modeling of RS devices with multiple finite states is presented. The proposed framework utilizes the finite-state Master Equations for Markov jump processes \cite{toral2014stochastic} for the development of compact model of the stochastic RS in such devices, encapsulating local irregularities of the stochastic state evolution and multilevel programming, which are essential for the design of real applications such as high-radix arithmetic operations and neuromorphic computing. As a generic framework, the proposed approach is not focused to any particular type of RS devices, but aims to formulate a probabilistic basis for RS devices that suffer from stochastic switching. In this work, we expose the proposed theoretical principles to deliver different modeling paradigms, one for two-state and one for four-state (2-bit), respectively adopting characteristics of existing RS devices.

\section{Markov Jump Process Master Equations}
\label{sec:introduction}
The notion of master equations is widely used in various scientific fields, from classical mechanics and chemistry to quantum mechanics, where stochasticity is inherent property of the system, like particle kinetics, chemical reactions and quantum state vectors \cite{kaniadakis1993kinetic,pilling2003master,rivas2010markovian}, as they provide simplified mathematical formulation of complex stochastic systems. The master equations are sets of low-order differential equations that describe the evolution of the probability over time of a discrete-state system and can be solved without the use of nontrivial numerical methods alike the stochastic differential equations.

Let $S$ denote a system which, at any time $t$, can be in one of the states of $X={x_1,x_2,...,x_N}$ such $S(t)={x_i}$, where $i=1,2,...,N$ and $N\in\mathbb{N}$. We define the probability of $S(t)$ to be in state $x_i$ as $P_i(t)$. Assuming a Markov jump process with transition rate $w_{i,j}\geq0$ from $x_i$ to $x_j$ with $\{j=1,2,...,N~|~j\neq{}i\}$, the master equations that describe the time evolution of probability $P_i(t)$ read:

\begin{equation}
    \frac{dP_i(t)}{dt} = \sum_{\substack{j=1,i\neq{}j}}^{N} \Big[w_{j,i}\cdot P_j(t) - w_{i,j}\cdot P_i(t) \Big].
    \label{eq:dP_dt}
\end{equation}

\noindent In (\ref{eq:dP_dt}), right side, the first term derives from the finite Kolmogorov forward equations that correspond to the escape rates between states, whereas the second term derives from the finite Kolmogorov backward equations where the probability to stay in $x_i$ is expressed as the negative sum of the transition rates from all $x_j$ to $x_i$. Moreover, the conservation of total probability holds, i.e.:

\begin{equation}
    \frac{d}{dt} \sum_{i=1}^{N} P_i(t) = 0
    \label{eq:conservation}
\end{equation}

\noindent resulting in $\sum_{i=1}^{N} P_i(t) = 1$, $\forall t$, if the initial state of the systems is $\sum_{i=1}^{N} P_i(t_0) = 1$. Thus, equation (\ref{eq:dP_dt}) stands for any value of $w_{i,j}$, which are natural entities not bounded by the limitation of probabilities in the interval $[0, 1]$.

\section{RS Device Master Equations}
\label{sec:RSdevs}
By definition, RS devices are elements with memory, mathematically described by a state variable $x$ and with switching behaviour attributed to various physical entities, e.g. the length of the conductive filament. Even if the direct correlation of such devices and memristor notion is still arguable \cite{kim2019experimental}, the mathematical description of memristor is often used to develop RS devices' models. Thus, along with the state equation(s) that describe the state evolution over time, the conductance equation of the RS device is formulated by the state-dependent Ohm's law. A general form of a voltage-controlled memristor that models RS devices reads:

\begin{IEEEeqnarray}{rCl}
    i(t) &=& g(x(t),v(t))\label{eq:Volt_Statefull_Ohms_law}\\
    \frac{dx(t)}{dt} &=& f(x(t),v(t)) \label{eq:state_evolution}
\end{IEEEeqnarray}

\noindent where $i(t)$, $v(t)$, and $x(t)$ are the time-dependent current through, voltage across and state of the device, respectively. Moreover, $g(x(t),v(t))$ is a function of the device's conductance and $f(x(t),v(t))$ is the function that governs the state evolution of the device, both dependent on the state and voltage across the device, which represents the external stimuli in this voltage-dependent form. Accordingly, the current-dependent form of such systems can be extracted by interchanging $i(t)$ and $v(t)$ in (\ref{eq:Volt_Statefull_Ohms_law})-(\ref{eq:state_evolution}) and replacing $g(x(t),v(t))$ with a state- and current-dependent function of device's resistance $r(x(t),i(t))$. Without loss of generality, the voltage-dependent form is considered for the rest of the manuscript, where the externally applied voltage represents the system's input.

Focusing on the state evolution function $f(x(t),v(t))$, it is obvious that the state at any time $\tau = t+\Delta{}t$, where $t, \Delta{}t>0$ stands and $[t, t+\Delta{}t]$ is an infinitesimal interval, is calculated as a function of the state and the external stimuli at $t$. In particular, considering a discrete time evolution, which applies for any numerical integration method, the state at $\tau$ is $x_\tau = x_t + \Delta{}t\cdot{}f(x_t,v_t) = F(x_t,v_t,\Delta{}t)$, where $F$ is any discrete-time numerical integration method. Thus, it is clear that $x_\tau$ is independent of any previous state $x_s$, where $s<t$, and complies with the Markov \emph{memoryless} property\footnote{The memory of RS devices and the Markov property need not to be confused, as the first attributes the physical capability of the device to maintain its state under no external excitation.} that is required for a system to be described as a Markov process.

Assuming now a finite-state RS device with instantaneous switching between the states, the state evolution can be described as a Markov jumping process. This assumption stands for high transition speeds, which is valid for RS devices under high amplitude and/or low frequency stimuli; however, the proposed multilevel approach extends the validity of this assumption as it provides finer granularity between states.

Taking into account the above, the modeling of stochastic RS device can be performed by the combination of memristor's mathematical formulation and the master equations of Markov jump processes. The required step to model a RS device with the proposed mathematical framework is to estimate the transition rates $w_{i,j}$. Since the RS devices are dependent to the external stimuli and they are non-volatile, $w_{i,j}$ can be estimated by the combination of the voltage-dependent ($w^{v_t}_{i,j|t}$) and the state-dependent rates ($w^{x_t}_{i,j|t}$) that are also time-varying functions. The former one can be approximated by the switching frequencies between the states, which corresponds to the inverse of the switching times of the device, such as:

\begin{equation}
w^{v_t}_{i,j|t} = \big(t_{i,j}^{SW}(v_t)\big)^{-1},
\label{eq:trans_rate}
\end{equation}

\noindent where $t_{i,j}^{SW}(v_t)$ is the time needed by memristor  to switch from state $x_i$ to  $x_j$. These rates can be used directly in (\ref{eq:dP_dt}), denoting the voltage-dependent probability evolution equation, i.e. $dP^{v_t}_{i}/dt$. On the other hand, the state-conservation rates for state-dependent probability evolution, due to the non-volatility, are defined as:

\begin{IEEEeqnarray}{rCl}
    \frac{dP^{x_t}_{i,t}}{dt} &=& \sum_{\substack{j=1\\i\neq{}j}}^{N} \Big[-k(x_j,x_t) \cdot P_{i,t} + k(x_i,x_t) \cdot P_{j,t} \Big],\label{eq:conserv_rate}\\
    k(x_i,x_t) &=& \Bigg\{\begin{array}{cc}
     \eta_i, & x_i = x_t\\
     0, & x_i \neq x_t
  \end{array},\label{eq:k}
\end{IEEEeqnarray}

\noindent where $\eta_i$ is the conservation rate of state $x_i$ and (\ref{eq:conserv_rate}) satisfies (\ref{eq:conservation}). Thus, a finite-state RS device with probabilistic switching in discrete time is formulated as:

\begin{IEEEeqnarray}{rCl}
    i_t &=& g(x_t,v_t),\label{eq:Volt_Statefull_Ohms_law_discrete}\\
    \frac{d\mathbf{P}_t}{dt} &=& \frac{d\mathbf{P}^{v_t}_{t}}{dt} + \frac{d\mathbf{P}^{x_t}_{t}}{dt}\nonumber\\
     &=&  \mathbf{W}^{v_t}_t\cdot\mathbf{P}_t + \mathbf{W}^{x_t}_t\cdot\mathbf{P}_t =
    \mathbf{W}_t\cdot\mathbf{P}_t, \label{eq:prob_evolution}
\end{IEEEeqnarray}

\noindent where in bold are the vector-form of the probabilities  with size of $N\times{}1$, whereas, $\mathbf{W}_t$ is the transition matrix of the system composed of the voltage-dependent transition matrix $\mathbf{W}^{v_t}_t$ and the state-dependent conservation matrix $\mathbf{W}^{x_t}_t$, with size $N\times{}N$. In addition, the calculation of the state becomes implicit as it depends on the probability to be in any state and a random variable $U_t$, which follows a uniform distribution bounded in the interval [0, 1] and acts as the selector of the state, and it reads:

\begin{equation}
    \begin{array}{cc}
    x_t = x_i,& U_{t-1} \in \Big[\sum_{j=1}^{i-1}P_{j,t-1},\sum_{j=1}^{i}P_{j,t-1}\Big)
    \end{array}
    \label{eq:state}
\end{equation}

\noindent with the exception that $P_{N,t-1}$ includes the upper bound of the interval, i.e. the value $1$. Moreover, when a jump from $x_i$ to $x_j$ occurs with regard to (\ref{eq:state}), the probability vector $\mathbf{P}_t$ is redefined as $P_j,t=1$ and $P_l=0$, where $\{l=1,2,...,N~|~ l\neq{}j\}$, because the probability of the system to be at $x_j$ in the next time-step is the maximum.

\section{Validation}
The verification of versatility and expected functionality of the proposed stochastic RS device model is provided through different paradigms. More specifically, the most common case, namely a two-state (binary) RS device model, is formulated to show in details how the presented framework is applied. Furthermore, the multilevel capabilities of the framework are adequately delivered with a four-state (2-bit) probabilistic RS device example.

The proposed generic mathematical framework model covers the macro-modeling of the RS devices state evolution, so the conductance function $g(x_t,v_t)$ can be indifferent in regard to the conductance mechanisms that take place in the structure of each fabricated device, like for example quantum tunneling, Schottky emission,  Poole-Frenkel effect, etc. In particular, for the two selected examples, we will adopt the conductance mechanisms of previous works, and more specifically, ohmic conductance for the binary case and the combination of Schottky emission and ohmic conductance for the 2-bit case.

For the application of the aforementioned modeling framework, the transition matrix needs to be defined. In case of a simplified pulse programming scheme, where the amplitude of pulses is fixed, the transition rates can be pre-defined as a set of coefficients. However, for the general case of continuous range of input voltages, the voltage-dependent functions of transition rates require to be estimated.

So, starting from the binary example case ($N = 2$), measurements on fabricated RS devices with abrupt binary switching have shown that their switching time can be estimated by voltage-dependent log-normal or Poisson distributions \cite{medeiros2011lognormal,naous2015stochasticity}, according to the type of the RS device. Apparently, in this example, the Poisson-like switching from \cite{naous2015stochasticity} is adopted, i.e. $t^{SW}_{i,j|t} = \alpha / \text{exp}(v_t/\beta)$, where $\alpha$ and $\beta$ are fitting parameters. Thus, by using (\ref{eq:trans_rate}) and assuming:

\begin{equation}
    \mathbf{P}_t = \Bigg[\begin{array}{c}
     P_{on,t}\\
     P_{off,t}
  \end{array}\Bigg]
    \label{eq:p_2state}
\end{equation}

\noindent the transition matrix reads:

\begin{IEEEeqnarray}{rCl}
    \mathbf{W}_t &=& 
    \Bigg[\begin{array}{cc}
     -1/(t_{rst|t}^{SW}) & 1/(t_{set|t}^{SW}) \\
     1/(t_{rst|t}^{SW}) & -1/(t_{set|t}^{SW})
  \end{array}\Bigg]\nonumber\\
   &+&
  \Bigg[\begin{array}{cc}
     -k(R_{off}) & k(R_{on}) \\
     k(R_{off}) & -k(R_{on})
  \end{array}\Bigg] \label{eq:w_2state}
\end{IEEEeqnarray}

\noindent where $t_{rst|t}^{SW}$ ($t_{set|t}^{SW}$) is the voltage-dependent switching time for the transition $R_{on}$->$R_{off}$ ($R_{off}$->$R_{on}$), which is assumed infinite when $v_t\geq 0 $ V ($v_t\leq 0 $ V), and $k$ is calculated from (\ref{eq:conserv_rate}) with $\eta_{on}$ and $\eta_{off}$ representing the fitting parameters of the conservation rates.

\begin{figure}[!t]
\centering
  \includegraphics[width=0.80\linewidth]{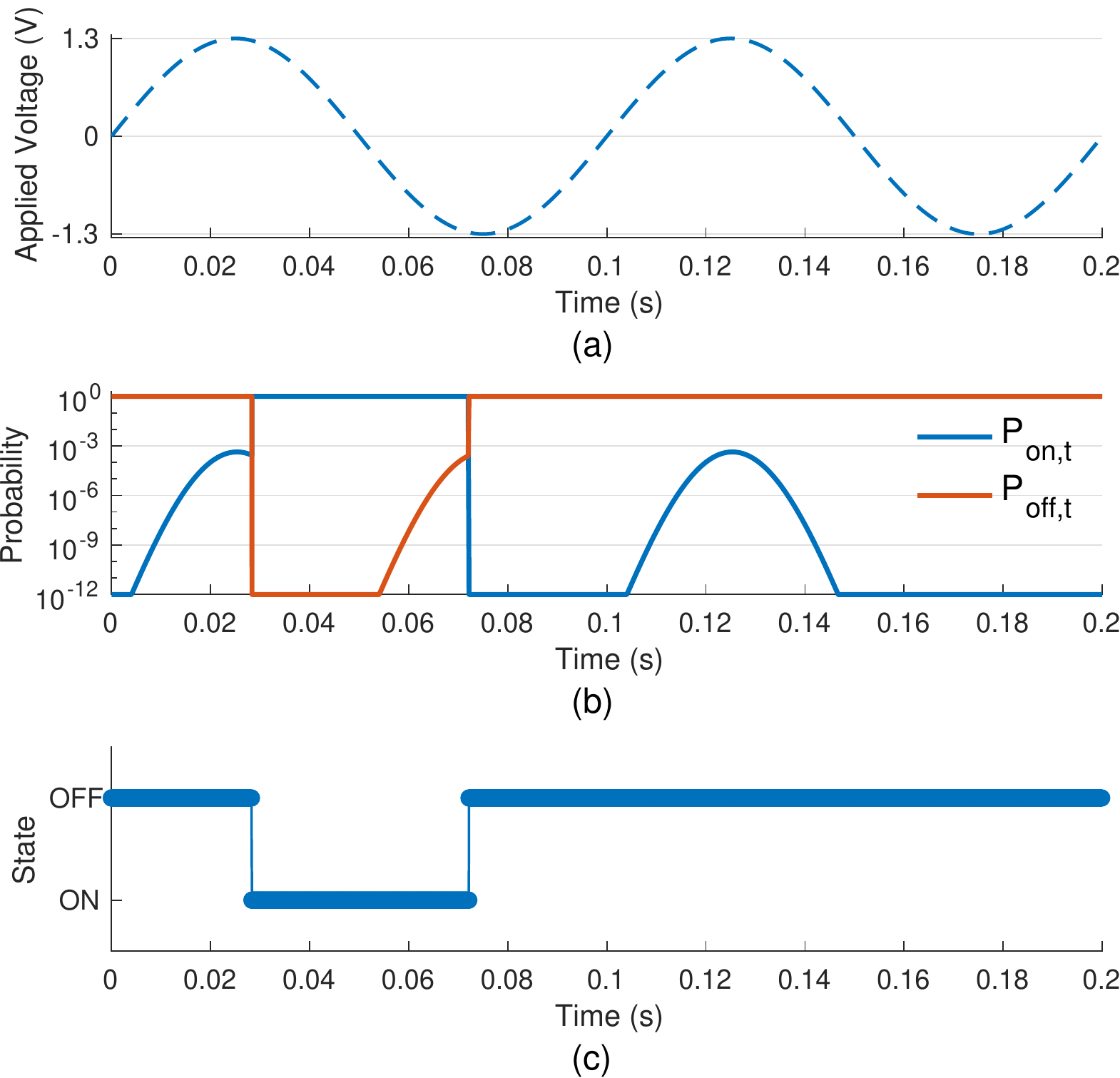}
    \caption{Binary stochastic switching. (a) Input voltage. (b) Probability and (c) State evolution over time. The simulation results obtained for $\alpha_{set}=\alpha_{rst}=3\times{}10^5$, $\beta_{set}=\beta_{rst}=0.05$, and $\eta_{on}=\eta_{off}=1.5\times{}10^6$.}
    \label{fig:2_state}
\end{figure}

\begin{figure}[!t]
\centering
  \includegraphics[width=0.80\linewidth]{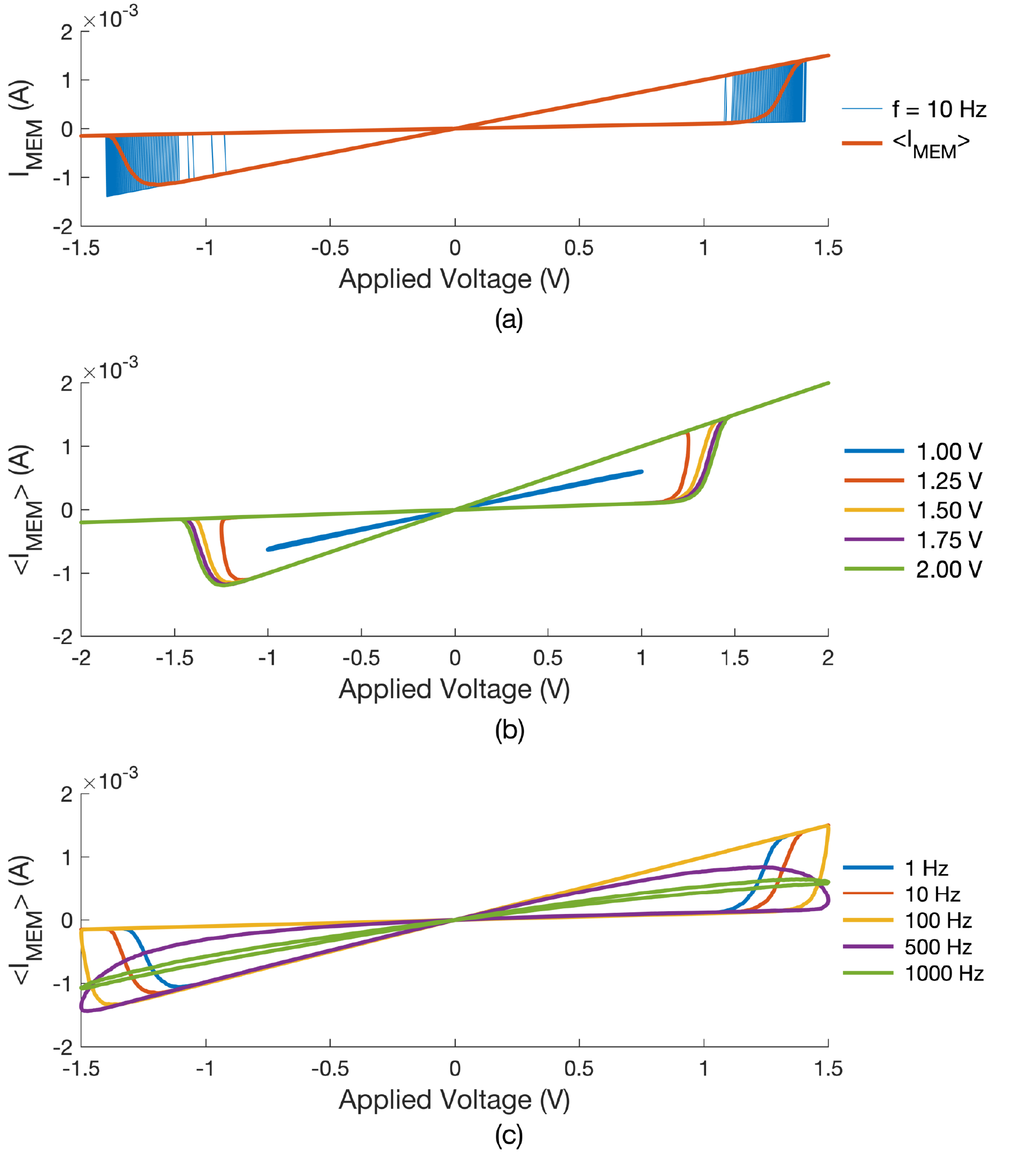}
    \caption{(a) I-V curve for binary stochastic switching and \textless$I_{MEM}$\textgreater. (b, c) \textless$I_{MEM}$\textgreater-V curve for various amplitudes and frequencies.}
    \label{fig:2_state_I-V}
\end{figure}

Utilizing (\ref{eq:Volt_Statefull_Ohms_law_discrete}), (\ref{eq:prob_evolution}), (\ref{eq:state}) with the probability vector (\ref{eq:p_2state}) and transition matrix (\ref{eq:w_2state}), the time evolution of the binary switching paradigm under a sinusoidal excitation is presented in Fig.~\ref{fig:2_state}. In particular, for the duration of two input signal periods, the probability and the state evolution are shown and for this specific realization, the jump $R_{off}$->$R_{on}$ never occurred for the second period, while the stochastic switching is also evident in the first period, where the two jumps occurred randomly in different time, even though the switching parameters are selected to be equal. A more general preview of the stochastic behavior of the binary model is illustrated through the I-V characteristic under sinusoidal excitation. Firstly, Fig.~\ref{fig:2_state_I-V}(a) shows a set of 1000 sweeps of fixed frequency and voltage amplitude $\{10$ Hz$, 1.5$ V$\}$ and the ensemble average of the current \textless{$I_{MEM}$}\textgreater{} (orange line), while Figs.~\ref{fig:2_state_I-V}(b, c) illustrate the \textless{$I_{MEM}$}\textgreater{} over 1000 sweeps for different voltage amplitude and frequency values, respectively. The frequency dependency is one of the key features of memristors.

Furthermore, another, multi-state this time paradigm, with a 2-bit ($N=4$) stochastic RS device model is presented. In this case, the transition rates are estimated according to the switching time of a deterministic multistate model \cite{garcia2016spice} that describes the behavior of fabricated multilevel RS devices such as in  \cite{chen2015switching}. Under these circumstances, the probability vector $\mathbf{P}_t$ is a $4\times{}1$ vector, while the size of the transition matrix $\mathbf{W}_t$ is $4\times{}4$ and its values are estimated from the energy-dependent switching in \cite{garcia2016spice}. In particular, the switching time between the states is expressed as function:
\begin{equation}
t^{SW}_{i,j|t} = \gamma_{i,j} /(v_t\cdot{}I_{i|t}),
\end{equation}

\noindent where $I_{i|t}=\zeta_i\cdot\text{exp}(\sqrt{|v_t|})$ for the states with Schottky emission ($i = 1, 2, 3$) and $I_{i|t}=\zeta_i\cdot{}v_t$ for the ohmic conductance state ($i = 4$). The value of $\zeta_i$ derives from an amount of parameters in \cite{garcia2016spice} and can be embodied in the parameter $\gamma_{i,j}$, so the voltage-dependent switching time reads:

\begin{equation}
    t^{SW}_{i,j|t} = \Bigg\{\begin{array}{cc}
     \gamma_{i,j} / \big(v_t\cdot\text{exp}(\sqrt{|v_t|})\big), & i = 1, 2, 3\\
     \gamma_{i,j} / v_t^2, & i = 4
  \end{array}.\label{eq:4_state_SW}
\end{equation}

\begin{figure}[!t]
\centering
  \includegraphics[width=0.80\linewidth]{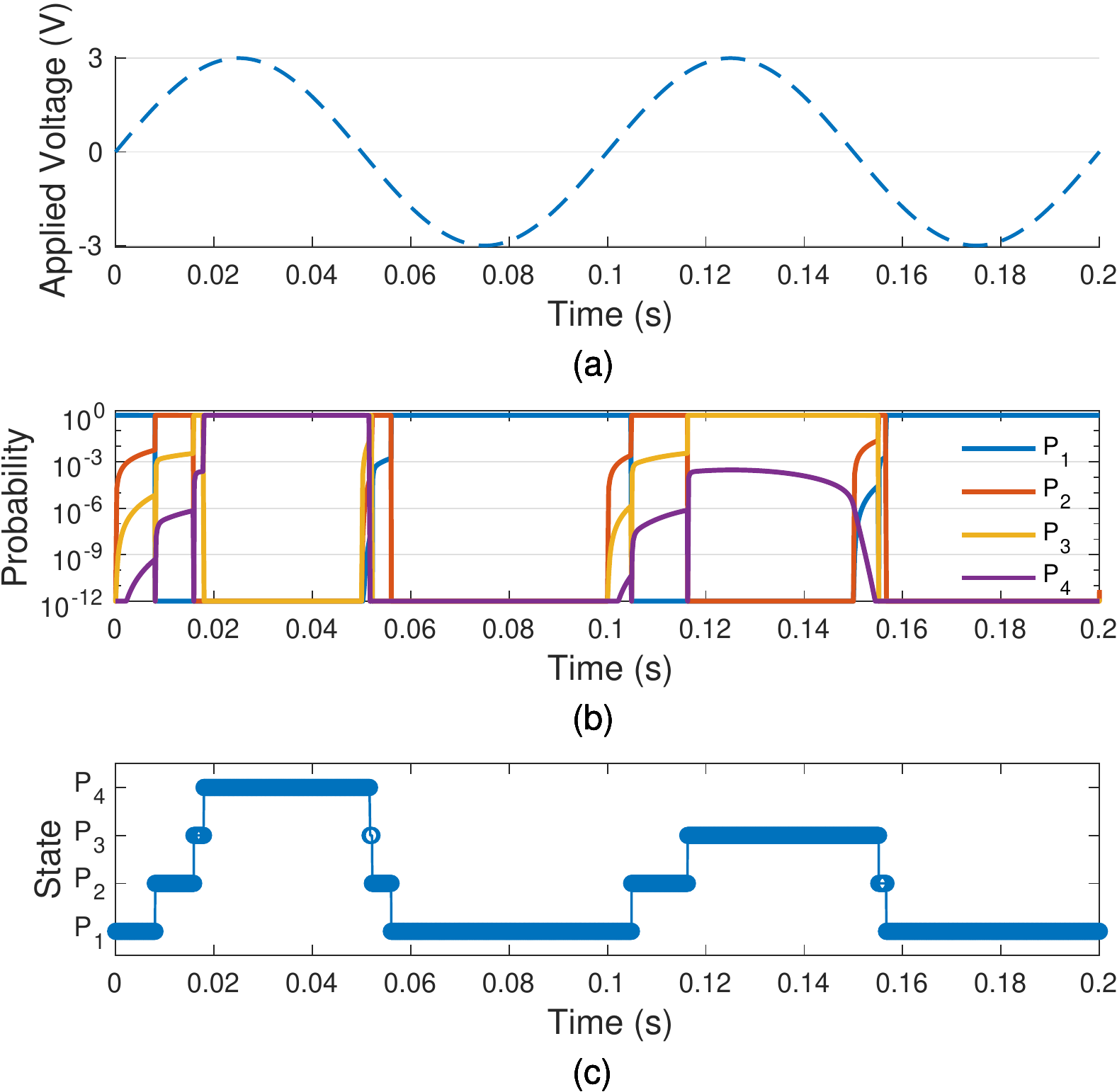}
    \caption{2-bit ($N=4$) stochastic switching. (a) Input voltage, (b) Probability and (c) State evolution over time. The simulation results were obtained for $\{\gamma_{1,2}, \gamma_{2,3}, \gamma_{3,4}, \gamma_{4,3}, \gamma_{3,2}, \gamma_{2,1}\}=\{0.263, 1.155, 19.11, 9.15\times{}10^{-4}, 3.06\times{}10^{-2}, 0.578\}$ and $\eta_{1}=\eta_{2}=\eta_{3}=\eta_{4}=3\times{}10^8$.}
    \label{fig:4_state_detailed}
\end{figure}

\begin{figure}[!t]
\centering
  \includegraphics[width=0.80\linewidth]{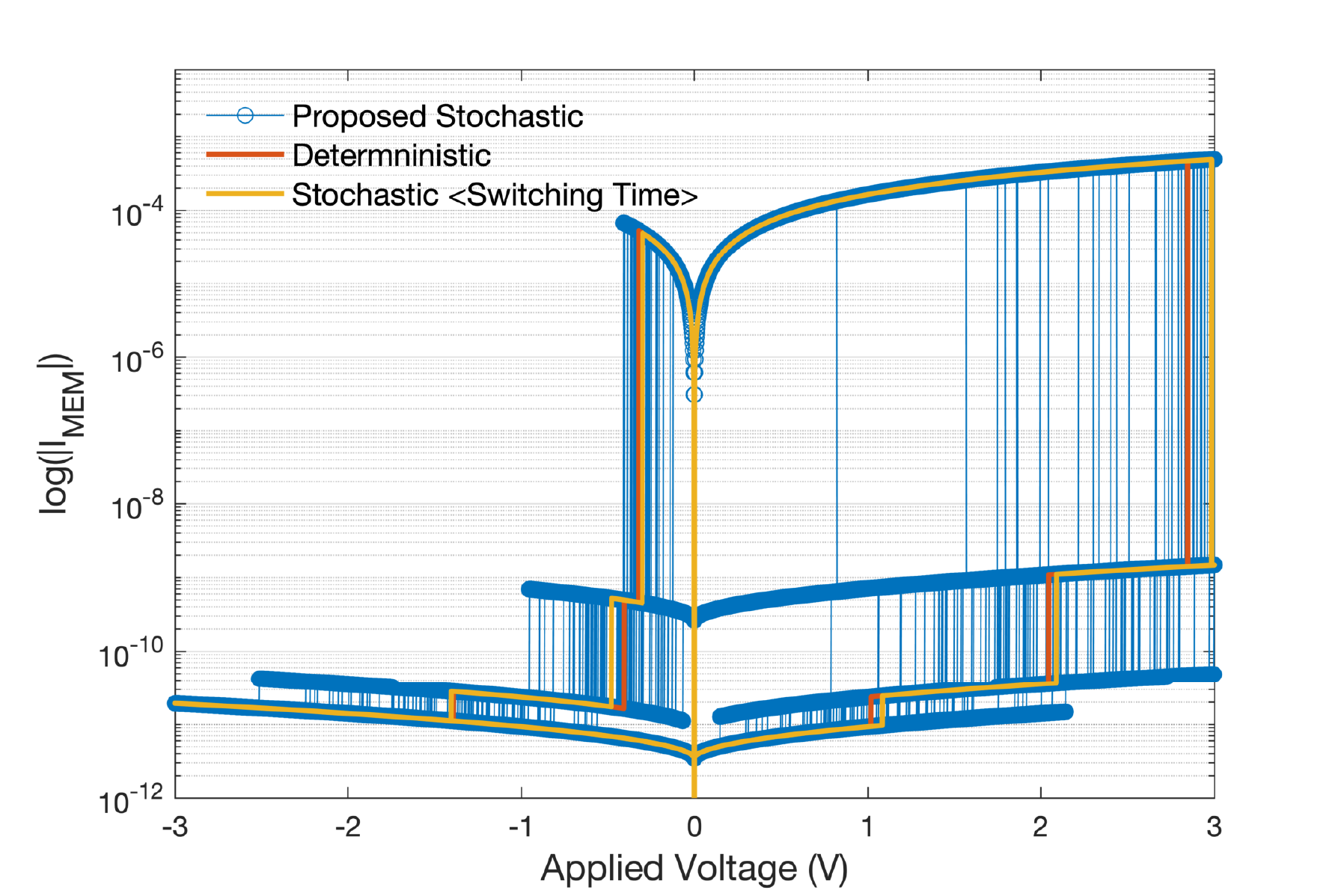}
    \caption{Logarithmic I-V curve of the 2-bit ($N=4$) paradigm along with the deterministic RS device model from \cite{garcia2016spice} and the ensemble mean switching time between states of the stochastic paradigm.}
    \label{fig:4_state}
\end{figure}

Equations (\ref{eq:4_state_SW}) and (\ref{eq:trans_rate}) are used to construct $\mathbf{W}^{v_t}_t$ for the $N=4$ paradigm, whereas (\ref{eq:conserv_rate}) and (\ref{eq:k}) with state conservation rates $\eta_i$  are used for $\mathbf{W}^{x_t}_t$. Similarly to the binary paradigm, the switching time $t^{SW}_{i,j|t}$ for $i>j$ ($j<i$) is assumed infinite when $v_t\geq 0 $ V ($v_t\leq 0 $ V), as well as when $|i-j|>1$. The input voltage, the probability of the 4 states and the state evolution over time for two randomly-selected periods of the applied voltage are presented respectively in Fig.~\ref{fig:4_state_detailed}. Under the influence of the applied voltage, the probability of being in each state is varying, while the stochastic switching is clearly depicted in Fig.~\ref{fig:4_state}(c) through the randomly occurrence of transitions between the states. In addition, Fig.~\ref{fig:4_state} presents the I-V characteristic of the 2-bit stochastic model for a sinusoidal excitation with frequency 10 Hz during 100 periods, in comparison to the deterministic 2-bit RS device model from \cite{garcia2016spice}, along with the ensemble mean switching time between states of the stochastic one (\textless{$Switching~Time$}\textgreater{}). The parameters were fitted according to the ensemble mean switching time. In Fig.~\ref{fig:4_state}, the four discrete conductance levels are illustrated, where the three lower levels correspond to low Schottky emission currents and the highest one to high ohmic current, as in \cite{garcia2016spice}. The stochastic switching between the states of the proposed RS device model is evident through the non-deterministic jumps depicted in the blue lines. The presented I-V results demonstrate the robustness of the proposed modeling approach to meet efficiently with other modeling approaches in both a qualitative and quantitative manner.

\section{Conclusions}
In this work, we postulate a novel probabilistic framework to model the stochastic resistive switching of nano-devices using multiple finite states, surpassing the limitations of the existing binary RS device stochastic models. Such a model has a direct application in multi-bit storage, multi-value logic operations and neuromorphic computing when limited number of resistive states is used due to the constrains of the state retention of fabricated RS devices. The mathematical background of the proposed framework is based on the formulation of memristor and the Master Equations of the Markov Jump processes, which can be both integrated to conventional circuit simulators through either SPICE or compact Verilog-A models. Next steps are the mitigation to any of these circuit simulation platforms, along with the introduction of higher-level features, such as the resistance range degradation due to aging or over-tuning in order to establish a complete probabilistic platform to evaluate the reliability of the RS device-based systems.

\section{Acknoledgement}
This work was supported by the Hellenic Foundation for Research and Innovation (H.F.R.I.) under the “First Call for H.F.R.I. Research Projects to support Faculty members and Researchers and the procurement of high-cost research equipment grant” (Project Number: 3830).

\bibliographystyle{IEEEtran}
\bibliography{ms.bib}

\newpage

\end{document}